\documentclass[10pt,journal,final]{IEEEtran}

\usepackage{cite}

\usepackage[T1]{fontenc} 
\usepackage{amsmath}
\usepackage{bm} 

\usepackage{stfloats}

\usepackage{graphicx}
\usepackage{epstopdf}
\usepackage{subfigure}

\usepackage{url}

\usepackage{booktabs}
\usepackage[binary-units]{siunitx}

\begin{document}

\title{Steady State Visually Evoked Potentials detection using a single 
electrode consumer-grade EEG device for BCI applications}

 
\author{Enrico Calore
\thanks{This work was conducted when E.~Calore was a PhD
student with the Department of Computer Science, Universit\`a degli Studi di 
Milano, Via Comelico 39/41, I-20135 Milan, Italy.}
\thanks{E.~Calore is now a Research Assistant with Universit\`a degli Studi di 
Ferrara, Via Saragat 1, I-44122 Ferrara, Italy. (e-mail: enrico.calore@unife.it)}
}
 
\IEEEspecialpapernotice{(This work was conducted between 2013 and 2014)}
  
\maketitle

\begin{abstract}
Brain-Computer Interfaces (BCIs) implement a direct communication pathway
between the brain of an user and an external device, as a computer or a 
machine in general.
One of the most used brain responses to implement non-invasive BCIs is the so 
called steady-state visually evoked potential (SSVEP).
This periodic response is generated when an user gazes to a light flickering 
at a constant frequency.
The SSVEP response can be detected in the user's electroencephalogram (EEG) at 
the corresponding frequency of the attended flickering stimulus.
In SSVEP based BCIs, multiple stimuli, flickering at different frequencies, are 
commonly presented to the user, where to each stimulus is associated
a command for an actuator.
One of the limitations to a wider adoption of BCIs is given by the need of EEG 
acquisition devices and software tools which are commonly not meant for end-user 
usage.
In this work, exploiting state-of-the-art software tools, the use of a low cost 
easy to wear single electrode EEG device is demonstrated to be exploitable to 
implement simple SSVEP based BCIs.
The obtained results, although less impressive than the ones obtainable with 
professional EEG equipment, are interesting in view of practical low cost BCI 
applications meant for end-users.
\end{abstract}


\section{Introduction}
\label{sec:ssvep-mindset}

Brain-Computer Interfaces (BCIs) implement a direct communication pathway
between the brain of an user and an external device, as a computer or a 
machine in general~\cite{wolpaw2002brain}.
The purpose of a BCI is to translate a detectable brain state of the user in a 
command for an actuator, providing the user with a real-time feedback.
BCIs may record the brain activity of their users with different methods, but 
the most used is electroencephalography (EEG).
EEG recording of the neural activity allow for non-invasive recordings with a 
relatively high time resolution and the use of relatively inexpensive devices, 
with respect to other methods. 

Initial research regarding BCIs aimed to provide mobility-impaired users
with a tool capable of translating a thought or a will into a command for an
external device or a prosthetic limb. 
Nevertheless, more recently, BCIs are gaining attention also as new means to 
interact with computers and other devices for healthy subjects 
too~\cite{bci-beyond-medical}.

In literature, different BCI modalities have been successfully adopted and
differentiate between them accordingly to the kind of underlying brain process
that is investigated to detect features associated to brain states.
The most popular are the P300 or Event Related Potentials (ERP), Motor Imagery
(MI) or Event Related Synchronization/De-synchronization (ERS/ERDS) and
Steady-State Visually Evoked Potentials.

BCIs could be further divided into three main categories, with
smooth boundaries~\cite{zander2010enhancing}; every modality may fit inside
one of these category, accordingly to how it is used, although not every
modality is suitable for every category:

\begin{itemize}
 \item \textit{Active} BCIs derive their outputs from brain activity 
which is
directly consciously controlled by the user, independently from external events,
for controlling an application.
 
 \item \textit{Reactive} BCIs derive their outputs from brain activity arising
in reaction to external stimulation, which is indirectly modulated by the user
for controlling an application.
 
 \item \textit{Passive} BCIs derive their outputs from arbitrary brain activity
without the purpose of voluntary control, for enriching a human-computer
interaction with implicit information.
\end{itemize}

The performance of a consciously controlled BCI (\textit{Active} or 
\textit{Reactive}) is commonly given in terms of its Information Transfer 
Rate (ITR).
The ITR value has been introduced in order to take into account both the speed
of a BCI in detecting a user command and its accuracy in detecting the
correct command~\cite{mcfarland2003brain}.
BCI research is consequently strongly focused on the improvement of the 
detection accuracy and the time needed to issue a command. 
Both of these factors are strongly dependent on the acquired signals quality, 
as well as the efficiency of the feature extraction algorithms, machine 
learning methods and the potential training of the users.

Other performance metrics for a BCI could be its cost in terms of hardware, 
or its ease of use, concerning the time to wear the electrodes system and the 
training time needed for calibrations or for the user to learn to modify 
her/his brain activity to issue commands.

One of the limitations to a wider adoption of BCIs is given by the need of data 
acquisition devices which are commonly not meant for end-user usage, requiring 
trained technicians to use them and moreover being orders of magnitude more 
expensive than ordinary interaction devices.
Furthermore, only recently general purpose BCI software frameworks are being
developed in order to provide user friendly tools for non-programmers.

The aim of this work is indeed to exploit state-of-the-art software tools and 
feature extraction algorithms in order to investigate the feasibility of an 
easy to use BCI, minimizing the hardware cost and the user time needed for 
the system set up.

In particular a low cost easy to wear single electrode EEG device is 
investigated in order to implement a \textit{Reactive} BCI based on the SSVEP 
response detection.

\section{SSVEP based BCIs}

In this research work the SSVEP modality, commonly used to implement 
\textit{Reactive} BCIs, has been chosen because of its high reachable ITR, 
the short calibration time needed, the low number of EEG electrodes 
required and also for the low BCI illiteracy showed~\cite{ssvep-illit}, 
granting high usability for most of the users, also in out-of-laboratory 
environments.
In fact, in the last years the SSVEP response has been adopted widely for 
the implementation of \textit{Reactive} BCIs~\cite{vialatte2010steady}.

Steady-state VEP (SSVEP) are a particular case of Visually Evoked Potentials 
(VEP), where the visual stimulus is presented several times at a frequency at 
least higher than 3.5Hz, but more commonly higher than 6Hz.
In this case, a quasi-sinusoidal periodic response called SSVEP can be observed 
in the scalp recorded EEG signals, in particular over the occipital brain 
region, where the visual cortex resides.

The stimulus presented to elicit a SSVEP response is commonly a flickering
light and accordingly to the flickering frequency, in the EEG signals acquired
from an user looking at the stimulus an increase in the power at the 
corresponding frequency (and harmonics) can be detected.
Consequently in the case of SSVEP, the presence of the response can be detected 
analyzing the frequency spectrum of the recorded signal.

In presence of multiple stimuli, flickering at different frequencies in the
visual field of the user, the increase of the signal power in the same
frequencies of the stimuli is more pronounced in the single frequency
corresponding to the single stimulus the user is gazing at.

Thanks to this observation, \textit{Reactive} BCIs can be implemented using this
modality, since the user, gazing at a particular stimulus chosen from the
presented ones, can issue to the system a command previously associated to the 
stimulus.

For BCI applications, stimuli are commonly presented by LED lights, or by 
shapes on a regular computer monitor, flickering at frequencies ranging between 
6Hz and 40Hz.
The provided visual stimuli characteristics are known to be highly relevant 
concerning the SSVEP response amplitude, indeed their frequency stability as 
well as their size (in terms of user visual angle), color, duty cycle, 
modulation depth, etc. have to be carefully controlled to obtain a stronger 
response and thus an easier detection~\cite{molina-stimprop}.

Data acquisition is commonly performed placing electrodes over the visual 
cortex in the occipital region of the user's brain using multiple electrodes 
EEG devices.

The SSVEP signal processing could be very simple and lot of SSVEP based BCIs 
implemented so far simply filter the raw EEG signals with narrow band-pass 
filters, centered on the stimuli frequencies, to later estimate the signal 
power in the frequency regions corresponding to the different stimuli.
Features corresponding to the power in the different frequency bands can then 
be used to detect the gazed target using a previously trained classifier.

Anyhow, it has been demonstrated that better performances can be obtained using 
more sophisticated methods as: evaluating the signal power also on higher 
harmonics for each stimulation frequency~\cite{krusienski2008harmonic}; merging 
the information coming from different electrodes using spatial 
filters~\cite{spatial-filters,ola}; or using as features SNR indexes representing 
the ratio between the SSVEP response power and the stimulus-uncorrelated brain 
activity occurring in the same frequency band~\cite{ola}.

The adoption of sophisticated signal processing methods do not commonly 
introduce complexity from the user point of view, on the other side it commonly 
improve the BCI performance, thus, as will be highlighted in this work, it 
permits to obtain a fair performance also when using lower quality EEG signals.

\section{Towards practical BCIs}

Conductive gel based EEG electrodes represent the state-of-the-art in 
terms of signals acquisition quality~\cite{Mihajlovic201214}, despite of this, 
the montage of EEG cups adopting them commonly requires a trained technician 
and moreover users need to wash their hair after usage.
Just the fact that the user would need another person to mount the EEG headset
on her/him is a strong limitation for end-user applications, as is also the 
high cost of electroencephalographers with respect to other interaction devices.

Custom hardware is being investigated~\cite{zander2011dry} in order to provide 
easier to use devices by means of dry electrodes or salted-water based
electrodes and also industrial companies are working towards this direction (as
Emotiv, Neurosky, Biosemi, g.Tec, etc.).
To exploit easy to wear EEG devices is indeed one of the most challenging 
research directions towards practical BCIs in the last 
years~\cite{mihajlovic2013dry,emotiv-ssvep}.

Professional general purpose EEG headsets, implementing dry electrodes,
are already available (e.g. the g.Tec g.SAHARA system),
but from their cost and their complexity (e.g. the need to choose electrodes
positions, connect the electrodes to the amplifiers, etc.) is clear that they
are still meant to be used mainly by BCI and clinical researchers, but not 
end-users.

New kind of EEG devices, meant to improve the usability by non-trained 
end-users, recently appeared on the market; they are commonly characterized by 
a rigid structure where electrodes are fixed on a pre-determined position and 
could be worn by the user himself without the need of external help.
These devices commonly adopt dry electrodes or water-based electrode 
technologies in order to not slime the user's hair, renouncing to the higher 
signal quality given by the use of conductive gel, but improving users' comfort.

A famous commercial device of this kind is the Emotiv 
EPOC~\footnote{\url{http://www.emotiv.com/epoc/}} which is a salted-water based 
14-electrodes system, that for example has been recently successfully used to 
implement SSVEP based BCIs~\cite{emotiv-ssvep,emotiv}.

One of the less expensive devices of this kind is the Mindset single dry 
electrode system provided by Neurosky Inc. which looks like regular headphones, 
a part from an additional arm holding a dry EEG electrode to be positioned on 
the forehead.
It was designed to provide an additional interaction mean for generic computer 
application, e.g. computer games, implementing \textit{Active} BCIs where the 
user can learn, using a bio-feedback approach, how to control her/his own brain 
activity and thus the features extracted from it.
The features extracted from the raw EEG data are computed by a proprietary 
algorithm, which is known to compute indexes based on relative clinical band 
power ratios, but no more details were disclosed about it by its manufacturer.
Interestingly raw EEG data can be read-out from this device as well.

The Mindset is clearly not designed to detect SSVEP responses, but in view of 
more practical BCIs, it would be very interesting if the Mindset could be used 
also for SSVEP based BCIs, thanks to its extremely low cost and for its ease to 
be wear.
In particular, in contrast to the \textit{Active} BCI its manufacturer had in
mind, it would be very interesting to be able to use it for \textit{Reactive} 
BCIs based on the SSVEP modality, since it would avoid the need of subject 
training.
Applications where a generic naive user could just wear it in order to be able 
to use the BCI, would be possible.

Anyhow, to use the Mindset for this purpose various challenges has to be faced: 
having a single electrode it will not allow to exploit spatial filtering 
algorithms~\cite{spatial-filters}; having a dry electrode it will provide a 
noisier signal than gel or water based ones~\cite{Mihajlovic201214} and 
moreover, as already mentioned, wearing it as indicated by its manufacturer, the 
electrode would be positioned on the forehead (roughly at Fp1 with respect to 
the 10-20 positioning system) and thus very far from the visual cortex where 
SSVEP responses are more intense~\cite{vialatte2010steady}.

\section{Material and Methods}

\subsection{Stimuli presentation}

In the performed experiments, the flickering stimuli presentation was provided 
on a regular \SI{60}{\hertz} LCD computer monitor (HP LP2065) thanks to a 
custom developed C++/OpenGL software~\cite{tesi-mia}.
The used flickering stimulus was a squared white patch subtending 
\SI{5}{\degree} of the user visual angle.
To achieve an optimal stimuli presentation, the flickering frequency control 
was synchronized to the vertical screen refresh as suggested 
in~\cite{cecotti2010reliable}.

The duty cycle of the stimulus was tuned for best performances according 
to the results discussed in~\cite{duty-cycle} and was $50\%$ for the 
\SI{15}{\hertz} stimulation frequency and $40\%$ for the \SI{12}{\hertz} 
one (since in this case a $50\%$ duty cycle is not possible due to the integer 
odd number of frames given a \SI{60}{\hertz} screen refresh).

\subsection{Data acquisition}

The EEG signal acquisition was performed using the MindSet device already 
mentioned.
Its single electrode is designed to be positioned on the forehead, roughly
at the Fp1 position. 
It acquires the EEG signal band-pass filtered between 
\SI{3}{\hertz} and \SI{100}{\hertz} at a sampling rate of \SI{512}{\hertz}, 
digitizing it at \SI{12}{\bit}.
It can be connected to a computer for data acquisition using a Bluetooth
connection. 
It incorporates a notch filter to remove power-line artifacts and implements 
proprietary algorithms for further signal cleaning and feature extraction.
In addition to the single acquisition electrode, the MindSet has also three
other contacts positioned over the left ear of the subject, which are used as 
ground and reference electrodes.
In particular, a part from the raw filtered EEG signal, it provides also 
proprietary dimensionless features representing the power strength in the
clinical frequency bands and also two \SI{1}{\hertz} sampled signals called 
\textit{e-Sense Attention} and \textit{e-Sense Meditation} values.
The \textit{Attention} and \textit{Meditation} values are computed thanks to a
proprietary algorithm and very few information are available about their actual
meaning. In the manufacturer intention, the subject wearing this device should
be able to learn to control these two values in order to be able to use
\textit{Active} BCI applications after several hours of training.

Data acquisition and software triggering handling, was performed using the 
OpenVibe software framework~\cite{openvibe}.
Interestingly the MindSet has an automatic on-line check to detect the contact
quality between the skin and the electrodes, that, instead 
of an impedance value as is commonly used, returns a SNR between what a 
proprietary algorithm considers as the EEG signal and what it identifies as 
artifacts.
This value can be acquired by OpenVibe too and displayed in real-time to the 
user while wearing the device.


\begin{figure}[!t]
\centering
\includegraphics[width=0.4\columnwidth]{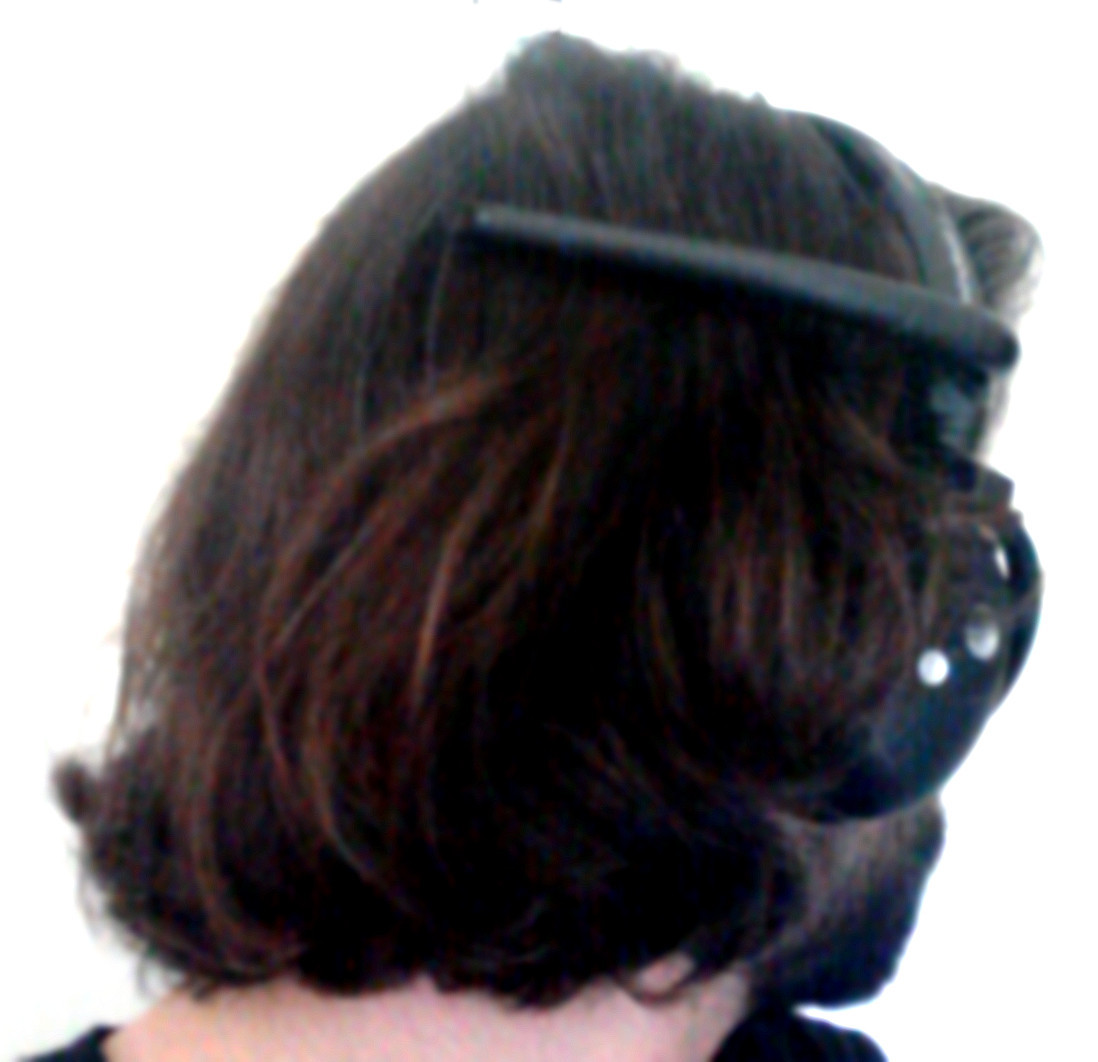}
\caption[Mindset reversed-wear]{Figure depicting an user wearing the Neurosky 
MindSet device ``reversed'', with its electrode backward facing, positioned 
roughly at P2 location (according to the extended 10-20 system), over the 
parietal lobe of the cerebral cortex.
\label{fig:reversed-mindset}}
\end{figure}

To overcome the wrong positioning of the MindSet's electrode for SSVEP response
detection, experiments was conducted using the MindSet on the subjects head, 
wearing it ``reversed'', letting the single electrode to be positioned 
backward and the reference/ground electrodes over the right ear.
In this manner, as shown in Fig.~\ref{fig:reversed-mindset}, the electrode is 
roughly positioned near P2 (according to the extended 10-20 system), 
which is a much more suitable location to detect the SSVEP response, although 
not the optimal one~\cite{vialatte2010steady}.

Being the MindSet electrode meant to be positioned on the forehead, its shape
is not appropriate to have a connection to the scalp where hair is present.
To overcome this problem a droplet of conductive electrode gel was used to
improve the contact.
Although consequences in terms of impedance could not be assessed due to the
proprietary hardware (electrode-skin impedance could not be measured and the
amplifier input impedance is unknown), experimental results, as exposed in the
next section, confirm that this procedure improve the acquisition signal
quality.
In order to have a similar impedance also on ground and reference electrodes,
a very small amount of gel was positioned also on them.

Despite of the use of a little amount of conductive gel, wearing the MindSet 
remains much easier than wearing ordinary gel based EEG devices and it could be 
easily done by the subject himself with no need of external help.
Moreover, the small amount of gel to be used, do not force the subject to have
a shower right after the use of the device.

The user, as already mentioned, could use the automatic impedance checker, 
implemented in the device, in order to assess the quality of the electrodes 
contact, which was reported in real-time and displayed by the acquisition 
software.

Montage of the device was in the range of about \SIrange{0.5}{3}{\minute},
according to the hair volume of the user.

The study was conducted in accordance with the recommendations of the 
Declaration of Helsinki and before the experiment, all participants signed 
the informed consent.

\subsection{Signal Processing}

The features describing the SSVEP response intensity were computed using the $T$ 
test statistic described in~\cite{ola}, where one of the best performing spatial 
filtering methods~\cite{spatial-filters} known as Minimum Energy Combination 
was proposed.
Having the MindSet a single electrode, spatial filtering is not possible, but 
the same approach discussed in~\cite{ola} was used to separate the SSVEP 
response contribution from the stimulus-uncorrelated brain activity.

The acquired EEG signal can be modeled as shown in Eq.~\ref{eq:model}, adapted 
from the multi-channel signal model described in~\cite{ola}.

\begin{equation}
\label{eq:model}
s(t) = \sum_{k=1}^{N_{h}} a_{k} \sin(2\pi kft + \phi_{k}) + \sum_{j}
b_{j} z_{j}(t) + e(t)
\end{equation}

The first component of $s(t)$ is the actual SSVEP response of interest, 
which is characterized by a set of sinusoids with frequency $f$ and its $k$ 
harmonics, each of which has a specific amplitude $a_{k}$ and phase
$\phi_{k}$.
The second component of the model is a set of signals $z_{j}(t)$, scaled by the 
weighting factors $b_{j}$, which are unrelated to the SSVEP response and 
comprise concurrent brain activity and internal as external artifacts. These 
signals are present.
Eventually, the last component $e(t)$ is a measurement noise component.

In vector form, the model can be expressed as $\mathbf{s} = \mathbf{xa} + 
\mathbf{zb} + \mathbf{e}$ and an estimate of the stimulus-uncorrelated 
component can be obtained as shown in Eq.~\ref{eq:ssvep-remove}.

\begin{equation}
\label{eq:ssvep-remove}
\tilde{\mathbf{s}} = \mathbf{s} - \mathbf{x} (\mathbf{x}^\top \mathbf{x})^{-1}
\mathbf{x}^\top \mathbf{s} \approx \mathbf{zb} + \mathbf{e}
\end{equation}

\noindent
Given this model, the actual $T$ statistic to be used as a feature, can be 
computed as shown in Eq.~\ref{eq:T}:

\begin{equation}
\label{eq:T}
T = \frac{1}{N_h} \, \sum_{k=1}^{N_h}
\frac{\hat{P}_{k}}{\hat{\sigma}_{k}^2}
\end{equation}

\noindent
where $\hat{P}_{k}$ is the estimated SSVEP power for the $k$-th harmonic
frequency in channel signal $\mathbf{s}$ and $\hat{\sigma}_{k}^2$ is an
estimate of the noise and uncorrelated brain activity in the same frequency. 
In other words, the $T$ statistic estimates how many time larger is the SSVEP
response power compared to the case where no visual stimulus is present,
averaging the SNRs ratios across $N_h$ harmonics.

The power $\hat{P}_{k}$ in the $k$-th harmonic is estimated as 
$\hat{P}_{k} = \| \mathbf{x}^\top_{k} \mathbf{s} \|^2$,
while, in order to avoid the need of calibration data acquired with no stimuli
presentation and also to take into account the nonstationarity of the noise, the
noise power $\hat{\sigma}_{k,l}^2$ is estimated on the same data segment, 
containing the SSVEP response, used to compute $\hat{P}_{k}$.

The SSVEP contribution is therefore removed from the signal as shown in 
Eq.~\ref{eq:ssvep-remove} to later fit an auto-regressive model $AR(p)$ of
order $p$ and use the fitted models to interpolate the noise power in the SSVEP 
frequencies.
The $AR(p)$ models are fitted using the Wiener-Khinchin theorem for computing
the autocovariance of the signal and then solving the Yule-Walker
equations using a Levinson-Durbin recursion~\cite{ola}.
This yields the $AR(p)$ parameters  $\alpha_1, \alpha_2, \dots, \alpha_p$ as
well as an estimate of the variance $\hat{\sigma}^2$ of the white noise driving
the $AR(p)$ process.
Once fitted the model to the signal $\mathbf{s}$, the noise level estimated at 
the $k$-th harmonic is given by:

\begin{equation}
 \hat{\sigma}_{k}^2 = \frac{\pi N_t}{4} \frac{\hat{\sigma}^2}{| 1 +
\sum_{j=1}^{p} \alpha_j \exp{(-2 \pi ijkf/F_s}) |^2}
\end{equation}

\noindent
where $N_t$ are the signal samples, $k$ is the harmonic number, $f$ is the 
stimulation frequency in \si{\hertz}, $F_s$ is the sampling frequency in
\si{\hertz} and $i = \sqrt{-1}$.

\section{Performed experiments}

\begin{figure}[!t]
\centering
\includegraphics[width=\columnwidth]{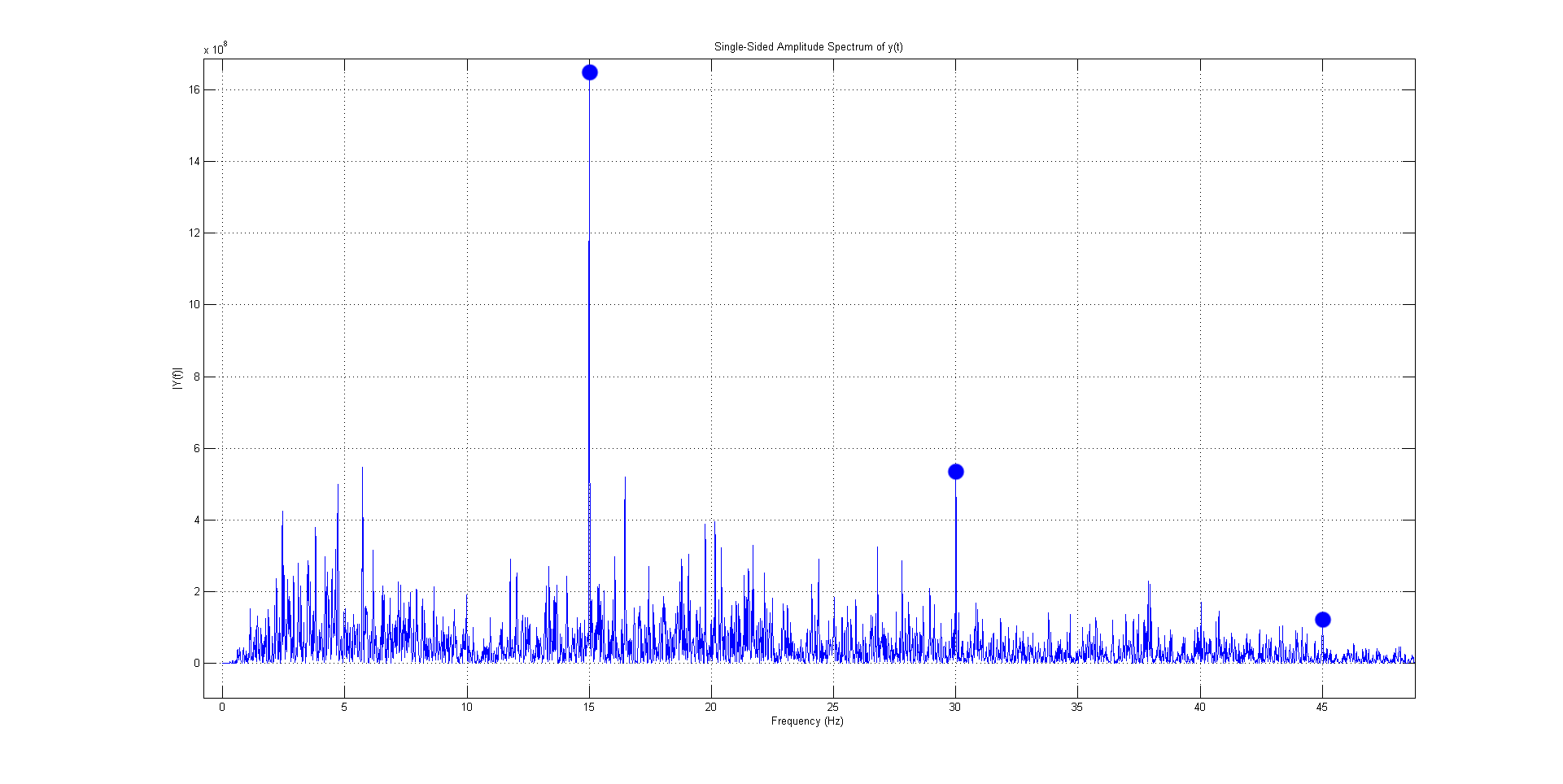}
\caption[Mindset SSVEP response detail]{
The PSD spectrum of the EEG signal acquired by the Mindset device positioned on 
the scalp of a subject gazing at a \SI{15}{\hertz} flickering white patch 
on a regular \SI{60}{\hertz} screen for \SI{30}{\second}.
\label{fig:mindset-ssvep2}}
\end{figure}

As a preliminary test to assess the possibility to record a SSVEP response 
using the MindSet unconventionally wore, as shown in 
Fig.~\ref{fig:reversed-mindset}, one subject was visually stimulated 
with a \SI{15}{\hertz} flickering white patch displayed on a screen and the PSD 
of its acquired EEG signal was computed off-line taking the FFT of the 
\SI{60}{\second} signal auto-correlation.
As shown in Fig.~\ref{fig:mindset-ssvep2}, the proposed setup led to a clear 
detection of a SSVEP response to the \SI{15}{\hertz} stimulation; peaks at the 
fundamental frequency and first harmonics are visible.

Knowing that the SSVEP response could be recorded with the MindSet, further
experiments have been conducted to identify the shortest signal length able to
lead to a classification accuracy high enough for BCI applications.
Having BCIs to be quasi real-time by definition, a SSVEP response detection 
has to be performed using at most few seconds long signal windows.
The ITR is indeed dependent on the accuracy, but on the detection speed as well.

In a similar fashion as performed in~\cite{ola}, multiple subjects were 
visually stimulated for \SI{30}{\second} for each trial for two different 
stimulation frequencies, chosen as \SI{12}{\hertz} and \SI{15}{\hertz}, while 
recording their EEG with the MindSet.
Four trials have been recorded for each subject in order to have a total of 
\SI{30}{\second} + \SI{30}{\second} recording for each stimulation frequency.

Off-line analysis was then performed, computing the $T$ test statistic reported 
in Eq.~\ref{eq:T} setting $N_h = 2$, for both the stimulation frequencies, for 
each signal window, irrespective of the actual stimulation frequency.
The same operation was performed for \SI{1}{\second} non-overlapping signal 
windows and then again, using \SI{2}{\second} non-overlapping signal windows.

Each signal window could be labeled with three values: 
$<f, T_{12}, T_{15}>$, the actual stimulation frequency $f$ and the two $T$ 
features representing the detected SSVEP response respectively for the two 
stimulation frequencies.

Information coming from the first trial for each subject and each stimulation 
frequency was used to train a simple linear least squares binary classifier, 
while information coming from the second trial was used to test it.
The datasets were not randomized between trials, since in an actual BCI 
application the training data would be acquired before the BCI use, so the same 
approach has been followed.

\section{Results}

\begin{figure}[!t]
\centering
\includegraphics[width=\columnwidth]{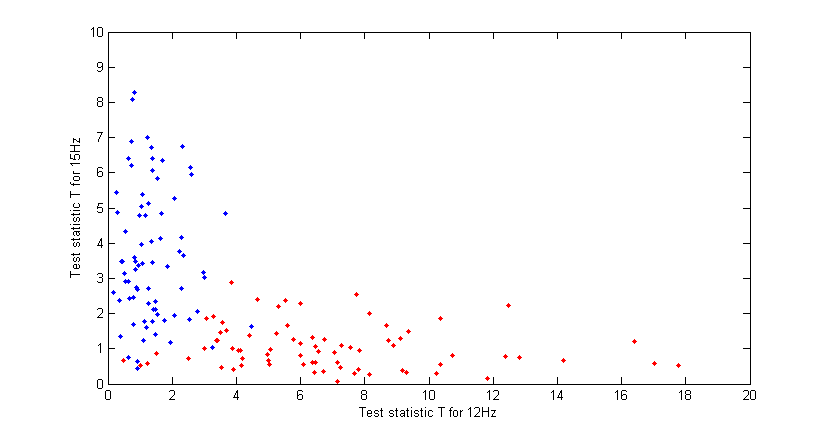}
\caption[Mindset test statistic $T$]{Test statistic $T$ computed for the two
frequencies for every two seconds non-overlapping window of EEG signal. Blu
points are epochs with a 15Hz stimulation, while red points are epochs with a
12Hz stimulation. Plotted data correspond to 10 trials acquired from one
subject, for a total of 150 epochs (75 for each frequency).
\label{fig:test-stat-t2}}
\end{figure}

Data points computed using two second signal windows, for one subject, are 
reported in Fig.~\ref{fig:test-stat-t2}, where the points color represents the 
actual stimulation frequency $f$ during the epoch, while the points coordinates 
are the computed features $(T_{12}, T_{15})$.

As can be seen, a linear classification between the epochs acquired under the 
two different stimulation frequencies seems to be feasible, despite of the 
short signal window (considering the used acquisition device).

\begin{figure}[!t]
\centering
\includegraphics[width=\columnwidth]{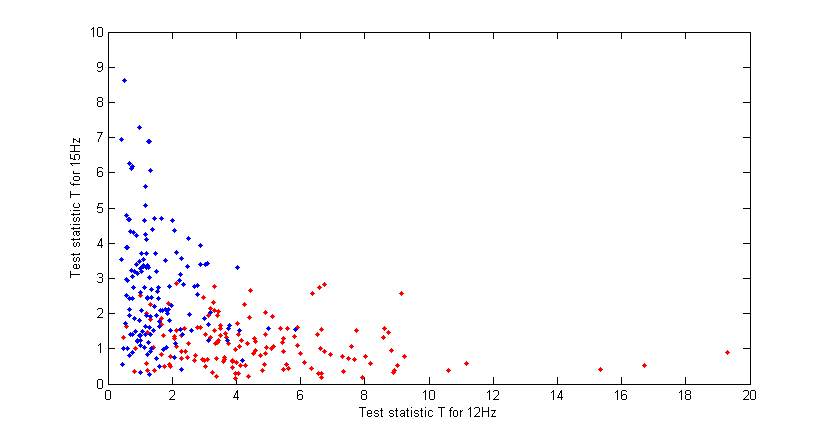}
\caption[Mindset test statistic $T$]{Test statistic $T$ computed for the two
frequencies for every one second non-overlapping window of EEG signal. Blu
points are epochs with a 15Hz stimulation, while red points are epochs with a
12Hz stimulation. Plotted data correspond to 10 trials acquired from one
subject, for a total of 300 epochs (150 for each frequency).
\label{fig:test-stat-t}}
\end{figure}

Using the same approach, the data points computed using one second signal 
window, on the same dataset are reported in Fig.~\ref{fig:test-stat-t}.

As can be seen, as expected using a shorter signal window, in this case, a 
linear classification would produce a lower accuracy, but nevertheless it seems 
to be feasible.

The classifier training was performed using the first trial acquired for each of
the two frequencies, while the remaining two trials where used as test sets;
results for the different subjects are reported in
Tab.~\ref{tab:mindset-results}.

\begin{table}[!t]
\renewcommand{\arraystretch}{1.3}
\caption{Classification accuracy. Results are reported for non-overlapping one 
second and two seconds signal windows.}
\label{tab:mindset-results}
\centering
\begin{tabular}{lcc}
\hline
     & \SI{1}{\second} windows & \SI{2}{\second} windows \\
\hline     
    Subject 1 & $90\%$ & $93\%$ \\ 
    Subject 2 & $83\%$ & $90\%$ \\ 
    Subject 3 & $74\%$ & $87\%$ \\ 
    Subject 4 & $70\%$ & $80\%$ \\ 
    Subject 5 & $69\%$ & $83\%$ \\ 
    Subject 6 & $50\%$ & $48\%$ \\ 
\hline

\end{tabular}
\end{table}

According to the reported results, for 5 subjects out of 6 the SSVEP response
can be detected with a reasonable accuracy.
As expected, using 2 seconds epochs lead to better results for all the subject
a part for the 6\textup{th} one.
For the 6\textup{th} subject, a manual inspection of the data points revealed
that the two point clouds, relative to the two stimulation frequencies, are not
separable for all the acquired trials.
The reason may be a SSVEP based BCI illiteracy of the subject, a very low 
attention payed to the flickering target or a particularly inefficient 
electrode location for the particular subject.

Concerning the classification accuracy, it is worth to mention that it has been
computed using non-overlapping windows, but, when implementing SSVEP based
BCIs, is a common practice to compute the SSVEP response index (e.g. the
frequencies power or the $T$ test statistic, as in this case) for sliding 
windows and then to evaluate the computed value for several subsequent windows.
This leads to a smoother output, lowering the effect of false-positive
detections which may be computed in a single signal window.
In this work this approach was not used to compute the values in
Tab.~\ref{tab:mindset-results}, since it would have been not a fair way to 
evaluate the classification accuracy, in the sense that multiple points would 
have been computed from the same parts of acquired signal.

\section{Conclusion and future works}

With the previously described experiment has been demonstrated that using a
popular single electrode consumer-grade EEG acquisition device is possible to 
detect a SSVEP response.
Moreover, despite of the not optimal electrode position and its physical shape, 
has been demonstrated that, using a state-of-the-art signal processing 
technique, the signal window length needed to accurately detect the SSVEP 
response could be short enough for BCI applications.

Moreover, in an actual SSVEP based BCI application adopting the proposed method, 
the real-time feedback, which was not presented in the performed experiments, as 
already demonstrated in other contexts~\cite{lotte2013combining}, should 
increase the users' SSVEP response intensity and thus enhance the detection 
accuracy and/or shorten the needed signal windows.

The reported results highlight the feasibility to implement a simple SSVEP based 
BCI using the MindSet device and the presented signal processing method.
This is interesting due to the wide diffusion and affordable cost of this device.

Even more interestingly, this work highlights the possibility to design new 
simple single electrode devices with a more suitable electrode position for the 
SSVEP detection and also adopting a specific electrode shape to let it be 
positioned where hair is present without the need of conductive gel.
  
\bibliographystyle{IEEEtran}
\bibliography{references}

\begin{thebibliography}{10}
\providecommand{\url}[1]{#1}
\csname url@samestyle\endcsname
\providecommand{\newblock}{\relax}
\providecommand{\bibinfo}[2]{#2}
\providecommand{\BIBentrySTDinterwordspacing}{\spaceskip=0pt\relax}
\providecommand{\BIBentryALTinterwordstretchfactor}{4}
\providecommand{\BIBentryALTinterwordspacing}{\spaceskip=\fontdimen2\font plus
\BIBentryALTinterwordstretchfactor\fontdimen3\font minus
  \fontdimen4\font\relax}
\providecommand{\BIBforeignlanguage}[2]{{%
\expandafter\ifx\csname l@#1\endcsname\relax
\typeout{** WARNING: IEEEtran.bst: No hyphenation pattern has been}%
\typeout{** loaded for the language `#1'. Using the pattern for}%
\typeout{** the default language instead.}%
\else
\language=\csname l@#1\endcsname
\fi
#2}}
\providecommand{\BIBdecl}{\relax}
\BIBdecl

\bibitem{wolpaw2002brain}
J.~R. Wolpaw, N.~Birbaumer, D.~J. McFarland, G.~Pfurtscheller, T.~M. Vaughan
  \emph{et~al.}, ``{Brain-computer interfaces for communication and control},''
  \emph{Clinical neurophysiology}, vol. 113, no.~6, pp. 767--791, 2002.

\bibitem{bci-beyond-medical}
J.~van Erp, F.~Lotte, and M.~Tangermann, ``{Brain-Computer Interfaces: Beyond
  Medical Applications},'' \emph{Computer}, vol.~45, no.~4, pp. 26--34, 2012.

\bibitem{zander2010enhancing}
T.~O. Zander, C.~Kothe, S.~Jatzev, and M.~Gaertner, \emph{{Enhancing
  human-computer interaction with input from active and passive brain-computer
  interfaces}}, ser. {Human-Computer Interaction Series}.\hskip 1em plus 0.5em
  minus 0.4em\relax Springer, 2010, pp. 181--199.

\bibitem{mcfarland2003brain}
D.~J. McFarland, W.~A. Sarnacki, J.~R. Wolpaw \emph{et~al.}, ``{Brain-computer
  interface (BCI) operation: optimizing information transfer rates},''
  \emph{Biological psychology}, vol.~63, no.~3, pp. 237--251, 2003.

\bibitem{ssvep-illit}
C.~Guger, B.~Z. Allison, B.~Grosswindhager, R.~Pr{\"u}ckl, C.~Hinterm{\"u}ller,
  C.~Kapeller, M.~Bruckner, G.~Krausz, and G.~Edlinger, ``{How many people
  could use an SSVEP BCI?}'' \emph{Frontiers in Neuroscience}, vol.~6, no. 169,
  2012.

\bibitem{vialatte2010steady}
F.-B. Vialatte, M.~Maurice, J.~Dauwels, and A.~Cichocki, ``{Steady-state
  visually evoked potentials: focus on essential paradigms and future
  perspectives},'' \emph{Progress in neurobiology}, vol.~90, no.~4, pp.
  418--438, 2010.

\bibitem{molina-stimprop}
J.~Bieger, G.~G. Molina, and D.~Zhu, ``{Effects of Stimulation Properties in
  Steady State Visual Evoked Potential Based Brain-Computer Interfaces},'' in
  \emph{{32nd Annual International Conference of the IEEE Engineering in
  Medicine and Biology Society}}, 2010.

\bibitem{krusienski2008harmonic}
D.~J. Krusienski and B.~Z. Allison, ``{Harmonic coupling of steady-state visual
  evoked potentials},'' in \emph{{Engineering in Medicine and Biology Society,
  2008. EMBS 2008. 30th Annual International Conference of the IEEE}}, 2008,
  pp. 5037--5040.

\bibitem{spatial-filters}
G.~Garcia-Molina and D.~Zhu, ``{Optimal spatial filtering for the steady state
  visual evoked potential: BCI application},'' in \emph{{Neural Engineering
  (NER), 2011 5th International IEEE/EMBS Conference on}}, 2011, pp. 156--160.

\bibitem{ola}
O.~Friman, I.~Volosyak, and A.~Graser, ``{Multiple Channel Detection of
  Steady-State Visual Evoked Potentials for Brain-Computer Interfaces},''
  \emph{Biomedical Engineering, IEEE Transactions on}, vol.~54, no.~4, pp.
  742--750, 2007.

\bibitem{Mihajlovic201214}
V.~Mihajlovi{\'c}, G.~Molina, and J.~Peuscher, ``{To what extent can dry and
  water-based EEG electrodes replace conductive gel ones?: A Steady State
  Visual Evoked Potential Brain-computer Interface Case Study},'' in
  \emph{{BIODEVICES 2012 - Proceedings of the International Conference on
  Biomedical Electronics and Devices}}, 2012, pp. 14--26.

\bibitem{zander2011dry}
T.~O. Zander, M.~Lehne, K.~Ihme, S.~Jatzev, J.~Correia, C.~Kothe, B.~Picht, and
  F.~Nijboer, ``{A dry EEG-system for scientific research and brain--computer
  interfaces},'' \emph{Frontiers in neuroscience}, vol.~5, 2011.

\bibitem{mihajlovic2013dry}
V.~Mihajlovi{\'c}, G.~Garcia-Molina, and J.~Peuscher, ``{Dry and Water-Based
  EEG Electrodes in SSVEP-Based BCI Applications},'' in \emph{{Biomedical
  Engineering Systems and Technologies}}.\hskip 1em plus 0.5em minus
  0.4em\relax Springer, 2013, pp. 23--40.

\bibitem{emotiv-ssvep}
N.~Chumerin, N.~Manyakov, M.~van Vliet, A.~Robben, A.~Combaz, and M.~{Van
  Hulle}, ``{Steady-State Visual Evoked Potential-Based Computer Gaming on a
  Consumer-Grade EEG Device},'' \emph{Computational Intelligence and AI in
  Games, IEEE Transactions on}, vol.~5, no.~2, pp. 100--110, 2013.

\bibitem{emotiv}
Y.~Liu, X.~Jiang, T.~Cao, F.~Wan, P.~U. Mak, P.-I. Mak, and M.~I. Vai,
  ``{Implementation of SSVEP based BCI with Emotiv EPOC},'' in \emph{{Virtual
  Environments Human-Computer Interfaces and Measurement Systems (VECIMS), 2012
  IEEE International Conference on}}.\hskip 1em plus 0.5em minus 0.4em\relax
  IEEE, 2012, pp. 34--37.

\bibitem{tesi-mia}
E.~Calore, ``{Towards SSVEP based Brain-Computer Interfaces for Virtual Reality
  environments explicit and implicit interaction},'' Ph.D. dissertation,
  Universit{\`a} degli Studi di Milano, 2014.

\bibitem{cecotti2010reliable}
H.~Cecotti, I.~Volosyak, A.~Graser \emph{et~al.}, ``{Reliable visual stimuli on
  LCD screens for SSVEP based BCI},'' in \emph{{In Proc. of the 18th European
  Signal Processing Conference (EUSIPCO-2010)}}, 2010.

\bibitem{duty-cycle}
G.~Huang, L.~Yao, D.~Zhang, and X.~Zhu, ``{Effect of duty cycle in different
  frequency domains on SSVEP based BCI: A preliminary study},'' in
  \emph{{Engineering in Medicine and Biology Society (EMBC), 2012 Annual
  International Conference of the IEEE}}, 2012, pp. 5923--5926.

\bibitem{openvibe}
Y.~Renard, F.~Lotte, G.~Gibert, M.~Congedo, E.~Maby, V.~Delannoy, O.~Bertrand,
  and A.~L{\'e}cuyer, ``{Openvibe: An open-source software platform to design,
  test, and use brain--computer interfaces in real and virtual environments},''
  \emph{Presence: Teleoper. Virtual Environ.}, vol.~19, no.~1, pp. 35--53, Feb.
  2010.

\bibitem{lotte2013combining}
F.~Lotte, J.~Faller, C.~Guger, Y.~Renard, G.~Pfurtscheller, A.~L{\'e}cuyer, and
  R.~Leeb, ``{Combining BCI with virtual reality: Towards new applications and
  improved BCI},'' in \emph{{Towards Practical Brain-Computer
  Interfaces}}.\hskip 1em plus 0.5em minus 0.4em\relax Springer, 2013, ch.~10,
  pp. 197--220.

\end{thebibliography}

\end{document}